\documentclass[twocolumn]{aastex6}

\shorttitle{Climate Modeling of a Potential ExoVenus}
\shortauthors{Stephen R. Kane et al.}

\begin{document}

\title{Climate Modeling of a Potential ExoVenus}

\author{
  Stephen R. Kane\altaffilmark{1},
  Alma Y. Ceja\altaffilmark{1},
  Michael J. Way\altaffilmark{2,3},
  Elisa V. Quintana\altaffilmark{4}
}
\altaffiltext{1}{Department of Earth Sciences, University of
  California, Riverside, CA 92521, USA}
\altaffiltext{2}{NASA Goddard Institute for Space Studies, New York,
  NY 10025, USA}
\altaffiltext{3}{Department of Physics and Astronomy, Uppsala
  University, Uppsala, SE-75120, Sweden}
\altaffiltext{4}{NASA Goddard Space Flight Center, Greenbelt, MD
  20771, USA}
\email{skane@ucr.edu}


\begin{abstract}

The planetary mass and radius sensitivity of exoplanet discovery
capabilities has reached into the terrestrial regime. The focus of
such investigations is to search within the Habitable Zone where a
modern Earth-like atmosphere may be a viable comparison. However, the
detection bias of the transit and radial velocity methods lies close
to the host star where the received flux at the planet may push the
atmosphere into a runaway greenhouse state. One such exoplanet
discovery, Kepler-1649b, receives a similar flux from its star as
modern Venus does from the Sun, and so was categorized as a possible
exoVenus. Here we discuss the planetary parameters of Kepler-1649b
with relation to Venus to establish its potential as a Venus
analog. We utilize the general circulation model ROCKE-3D to simulate
the evolution of the surface temperature of Kepler-1649b under various
assumptions, including relative atmospheric abundances. We show that
in all our simulations the atmospheric model rapidly diverges from
temperate surface conditions towards a runaway greenhouse with rapidly
escalating surface temperatures. We calculate transmission spectra for
the evolved atmosphere and discuss these spectra within the context of
the James Webb Space Telescope (JWST) Near-Infrared Spectrograph
(NIRSpec) capabilities. We thus demonstrate the detectability of the
key atmospheric signatures of possible runaway greenhouse transition
states and outline the future prospects of characterizing potential
Venus analogs.

\end{abstract}



\section{Introduction}
\label{intro}

Exoplanetary science lies at the threshold of atmospheric
characterization for terrestrial planets. The plethora of exoplanet
discoveries from the {\it Kepler} mission \citep{bor16} revealed that
the occurrence rate of planetary sizes increases toward smaller masses
\citep{fre13,how13,pet13} and that there are gaps in the radii
distribution that hearken to the effects of planet formation
mechanisms \citep{ful17}. The launch of the {\it Transiting Exoplanet
  Survey Satellite (TESS)} will likely find numerous transiting
terrestrial planets around bright host stars that are amenable to
transmission spectroscopy follow-up observations
\citep{ric15,sul15}. These planets will produce an unprecedented
sample of terrestrial atmospheres from which to study the demographics
of comparative planetology \citep{lec15}.

A primary focus of the exoplanet science community is to identify key
targets for follow-up observations with ground and space-based survey
missions. In particular, the James Webb Space Telescope (JWST) is
anticipated to become a major contributor to the measurement and study
of exoplanetary atmospheres \citep{gar06,bea18}. With such a large
sample of potential targets, it is necessary to apply criteria to
decide how valuable follow-up resources will be utilized
\citep{kem18}. One such criterion for narrowing the number of
follow-up targets is their position relative to the Habitable Zone
(HZ), which will allow for atmospheric characterization to focus on
planets that may have temperate surface environments
\citep{kas93,kop13a,kop14}. The methodology of selecting terrestrial
planets within the HZ was applied to the {\it Kepler} discoveries to
construct a catalog of HZ planets for further investigations
\citep{kan16}. Combining such HZ planet discoveries with statistical
techniques with respect to {\it Kepler} data sensitivity results in
estimates of HZ planet occurrence rates \citep{kop13b,dre15}.

However, exoplanet detections using the transit method are
intrinsically biased toward short orbital periods \citep{kan08},
resulting in a rarity of HZ exoplanet transit detections (except for
those planets orbiting very low-mass stars). The vast majority of
transit detections lie interior to the HZ where the incident flux on
the planet can be many times the solar flux. Thus, many of the
exoplanets that will be favorable targets for atmospheric
characterization are far more likely to be Venus analogs (exoVenuses)
rather than Earth analogs. The frequency of terrestrial planets that
may exhibit runaway greenhouse spectral features was quantified with
the Venus Zone (VZ) by \citet{kan14}, with occurrence rates of
$0.32^{+0.05}_{-0.07}$ and $0.45^{+0.06}_{-0.09}$ for M dwarfs and GK
dwarfs respectively. Several potential Venus analogs were identified,
such as Kepler-69b \citep{kan13} and Kepler-1649b \citep{ang17}. The
confirmation of a runaway greenhouse hypothesis for these planets, and
others like them, depends upon reliable transmission spectroscopy
analysis \citep{kan13,bea17}.

In order to determine the detectability of particular atmospheric
characteristics, we present the results of an exhaustive climate
simulation analysis based on the properties of the Kepler-1649
system. In Section~\ref{exovenus}, we describe the exoplanet
Kepler-1649b in detail, highlighting the properties that make it a
viable exoVenus candidate. Section~\ref{climate} details the full
range of our climate simulations and the results of that
analysis. From our climate models, we generate simulated transmission
spectra and identify key spectral features, presented in
Section~\ref{spectra}. Implications of our study are discussed in
Section~\ref{discussion}, and we provide concluding remarks in
Section~\ref{conclusions} including prospects for future observations.


\section{A Potential ExoVenus}
\label{exovenus}

Considering habitable conditions which would permit the existence of
life, Venus-type planets are of particular interest. Although Venus is
similar to the Earth in size, density, and composition, the divergence
of atmospheric conditions between Venus and Earth has rendered Venus
uninhabitable. Thus, studying Venus analogs can elucidate constraints
on habitability since such planets can serve as a template for runaway
greenhouse conditions that may be the most common scenario of
atmospheric evolution for Venus/Earth-size planets, due to the
irreversible nature of a runaway greenhouse
\citep{kas88,lec13}. Furthermore, there are still many features
regarding the interior, surface, and atmospheric evolution of Venus
that are currently unknown, and a deeper understanding of our sister
planet is essential to our interpretation of terrestrial exoplanet
data. The present state of Venus knowledge is summarized by
\citet{tay18}.

\begin{figure}
  \includegraphics[angle=270,width=8.2cm]{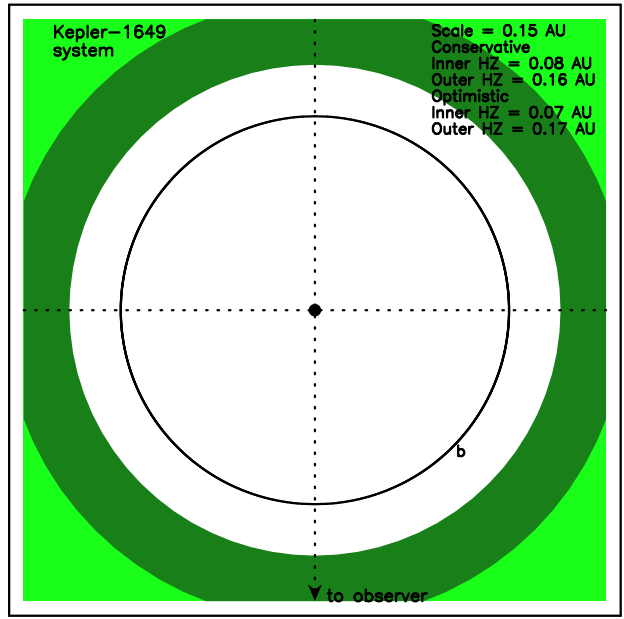}
  \caption{A top-down view of the Kepler-1649 system, showing the host
    star (intersection of the dotted cross-hairs), orbit of the `b'
    planet (solid line), conservative HZ (light-green), and optimistic
    HZ (dark-green). The scale of the figure is 0.15~AU along one edge
    of the box.}
  \label{hzfig}
\end{figure}

One of the most prominent potential Venus analogs discovered from {\it
  Kepler} observations is Kepler-1649b. The planet has a measured
radius of $R_p = 1.08 \pm 0.15$~$R_\oplus$ and receives a flux of $F_p
= 2.30 \pm 0.65$~$F_\oplus$ from its M dwarf host star \citep{ang17},
placing the planet firmly within the VZ of the system. Shown in
Figure~\ref{hzfig} is a top-down view of the Kepler-1649 system,
indicating the relative locations of the planetary orbit (solid line)
with respect to the star (intersection of the dotted cross-hairs). The
locations of the HZ regions are shown as light-green for the
`conservative' HZ and dark-green for the `optimistic' extension to the
HZ \citep{kop13a,kan16}. The inner edge of the optimistic HZ is based
on the assumption that Venus could have had surface liquid water as
recently as 1 Gya. However, because the surface record has been
substantially erased via resurfacing events \citep{tay18}, it is not
possible to confirm this hypothesis with currently available in-situ
data. Since Kepler-1649b is both similar in size and incident flux to
present-day Venus (1.9 versus 2.3 times present-day Earth insolation),
it may be considered an archetype of an exoVenus candidate. However,
like all VZ and HZ exoplanets, deductions regarding true surface
conditions are theories to be tested through climate modeling and
further observation.


\section{Climate Modeling and Surface Temperature}
\label{climate}

We attempt to model the atmosphere of Kepler-1649b using an atmosphere
with modern Earth constituents quite removed from that of modern
Venus. In fact, modeling the climate of Venus is non-trivial due to
the dramatic change in atmospheric chemistry that occurs at
atmospheric pressures above $\sim$10~bar and the lack of
high-temperature line lists for a large range of atomic and molecular
species \citep[e.g.][]{rot13}. Numerous climate models have been
adapted to the modern Venusian environment in an attempt to model its
dynamics and the temperature-pressure variations
\citep{leb10,and16,leb18}. Recently, similar such models have been
applied to a young, potentially habitable, Venus environment
\citep[][hereafter Way16]{way16}. Here we utilize the climate package
`Resolving Orbital and Climate Keys of Earth and Extraterrestrial
Environments with Dynamics' (ROCKE-3D), described in detail by
\citet{way17b}. ROCKE-3D has been used to model a variety of
terrestrial planet scenarios, such as synchronous rotation
\citep{fuj17} and the effects of variable eccentricity \citep{way17a}.

A large number of our initial simulations failed because the limit of
our radiation tables was rapidly reached. The upper limit is
approximately 400~K because we use the HITRAN 2012 database as
discussed below. We tested a total of 27 different scenarios in an
attempt to reach model stability as close as possible to estimates of
Kepler-1649b's insolation and rotation rate. To that end we slowed the
orbital period ($P$), lowered the incident flux ($F_p$), decreased the
initial CO$_2$ and CH$_4$ atmospheric composition, and adjusted the
topography and initial surface water. For all scenarios, we assumed
that the planet is tidally locked. Tidal locking can have a profound
effect on the climate circulation
\citep{kit11,wor15,kol16,car18,lew18}, and likely occurs on shorter
timescales and for longer-period planets than previously determined
\citep{bar17}. Given the relatively short tidal locking timescale for
short-period planets, the assumption of tidal locking is sufficiently
justified. A list of 10 key selected simulations (from a total of 27),
and their final surface conditions is shown in
Table~\ref{simtab}. These simulations give a summary of our attempts
to bring the model to a state as close to Kepler-1649b as possible
with ROCKE-3D.

\begin{deluxetable*}{ccccccccccccc}
  \tablewidth{0pc}
  \tablecaption{\label{simtab} Parameters of ROCKE-3D simulations.}
  \tablehead{
    \colhead{ID} &
    \colhead{$P$} &
    \colhead{$F_p$} &
    \colhead{Topo$^a$} &
    \colhead{CO$_2$} &
    \colhead{CH$_4$} &
    \colhead{Orbits$^b$} &
    \colhead{$T_\mathrm{mean}$$^c$} &
    \colhead{$T_\mathrm{min}$} &
    \colhead{$T_\mathrm{max}$} &
    \colhead{Bal$^d$} &
    \colhead{H$_{2}$O$_\mathrm{strat}$$^e$} &
    \colhead{H$_{2}$O$_\mathrm{surf}$$^f$} \\
    \colhead{} &
    \colhead{(days)} &
    \colhead{($F_\oplus$)} &
    \colhead{} &
    \colhead{(ppmv)} &
    \colhead{(ppmv)} &
    \colhead{} &
    \colhead{($^\circ$C)} &
    \colhead{($^\circ$C)} &
    \colhead{($^\circ$C)} &
    \colhead{(W\ $\mathrm{m}^{-2}$)} &
    \colhead{($\mathrm{kg\ kg}^{-1}$)} &
    \colhead{($\mathrm{kg\ kg}^{-1}$)}
  }
\startdata
1 &  8.6 & 2.3  & 1 & 400 & 1 &  300 & 90.2 & 81.2 & 112.5 & 103 & 0.379 & 0.4845\\ 
2 &  8.6 & 2.3  & 3 & 400 & 1 &   66 & 84.3 & 42.4 & 212.4 & 214 & 0.099 & 0.1168\\ 
3 &  8.6 & 2.3  & 3 & 100 & 0 &  116 &128.9 & 67.3 & 285.4 & 137 & 0.271 & 0.3456\\ 
4 &  8.6 & 2.0  & 1 & 400 & 1 &  486 & 91.8 & 83.5 & 111.8 &  77 & 0.423 & 0.5214\\ 
5 &  8.6 & 1.8  & 1 & 400 & 1 &  634 & 90.8 & 83.1 & 110.2 &  63 & 0.434 & 0.5061\\ 
6 &  8.6 & 1.6  & 1 & 400 & 1 &  752 & 88.0 & 81.3 & 102.7 &  59 & 0.411 & 0.4562\\ 
7 &  8.6 & 1.4  & 1 & 400 & 1 & 2031 & 64.1 & 57.7 &  72.0 &  35 & 0.026 & 0.1623\\ 
8 &  8.6 & 1.0  & 2 & 400 & 1 & 2345 &  4.3 &-54.1 &  46.9 & -3  & 0.000078 & 0.0064\\ 
9 & 16.0 & 1.47 & 2 & 400 & 1 & 2055 & 58.3 & 33.5 &  91.7 &  9  & 0.038 & 0.1052\\ 
10& 50.0 & 1.4  & 1 & 376 & 0 & 6354 & 59.0 & 56.3 &  61.9 & 0.3 & 0.027 & 0.1226 
\enddata
\tablenotetext{a}{1 = Aquaplanet, 2 = Venus, 3 = Venus bath.}
\tablenotetext{b}{The number of orbits the model was able to
  complete.}
\tablenotetext{c}{$T_\mathrm{mean}$,$T_\mathrm{min}$, and
  $T_\mathrm{max}$ are the global mean surface temperature averaged
  over the last 10 orbits of the run.}
\tablenotetext{d}{Balance is the net radiative balance at the
  end-point of the run averaged over the last 10 orbits. ROCKE-3D
  normally strives for this number to be within $\pm$0.2 W m$^{-2}$,
  so none of the runs were able to reach thermal equilibrium.}
\tablenotetext{e}{H$_{2}$O$_\mathrm{strat}$ is the specific humidity
  in units of H$_{2}$O (kg) to air (kg) in the stratosphere of the
  model (at 100~mb in the model atmosphere). Except for simulation 8,
  all others are above 10$^{-3}$ which implies they are approaching or
  are within the moist-greenhouse limit \citep{kas88}.}
\tablenotetext{f}{H$_{2}$O$_\mathrm{surf}$ is the specific humidity in
  units of H$_2$O (kg) to air (kg) at the surface layer of the
  model. Except for simulation 8, all others are above 10\% which
  implies that the atmospheric mass has increased enough to begin to
  cause substantial errors in the dynamics of ROCKE-3D.}
\end{deluxetable*}

The `Orbits' column indicates the number of planetary orbits the
simulation survived before the climate dynamics became unstable due to
the radiation tables moving beyond the valid boundaries of the model.
The upper bounds of the ROCKE-3D radiation tables are set by the
HITRAN 2012 \citep{rot13} line list which becomes invalid above 400~K
for the gases used in this study. However, in practice, heating rate
errors between atmospheric layers will start to grow for temperatures
above $\sim$340~K. We use 12 long-wave spectral bands and 24
short-wave bands which allow higher precision than the default
SOCRATES\footnote{ROCKE-3D uses the SOCRATES radiation package.} Earth
specific ga7 spectral files \citep{Edwards1996,ES1996}. The spectral
files were specifically designed for the atmospheres of planets around
M~dwarfs with higher temperatures and high specific
humidities. Typically, once the 400~K limit is reached, the radiation
tables start to produce temperature values out of bounds that are
indentified by the dynamics code that calculate the heating rates
between atmospheric layers (derived from the calculated radiation
fluxes). The dynamics code has preset bounds in place that purposely
stop the model when these bounds are exceeded.

The mean surface temperatures of the models at the end of the
simulations (regardless of whether the bounds were exceeded) are shown
in the column labeled `$T_\mathrm{mean}$'. The topographical model
used in each simulation is indicated by the `Topo' column, where the
three different categories are defined as follows: aquaplanet (899~m
deep all ocean planet), Venus (modern Venus topography, described in
Way16), and Venus bath (paleo-Venus topography with all-ocean grid
cells set to 1360m in depth). All simulations utilize a Kepler-1649b
specific BT-Settl \citep{Allard2012} spectral distribution file with
$T_{\mathrm{eff}}$ = 3200~K, $\log g = 5$, and [Fe/H] = 0.

\begin{figure*}
  \begin{center}
    \includegraphics[angle=270,width=16.0cm]{f02a.ps} \vspace{0.4cm}\\
    \includegraphics[clip,width=16.0cm]{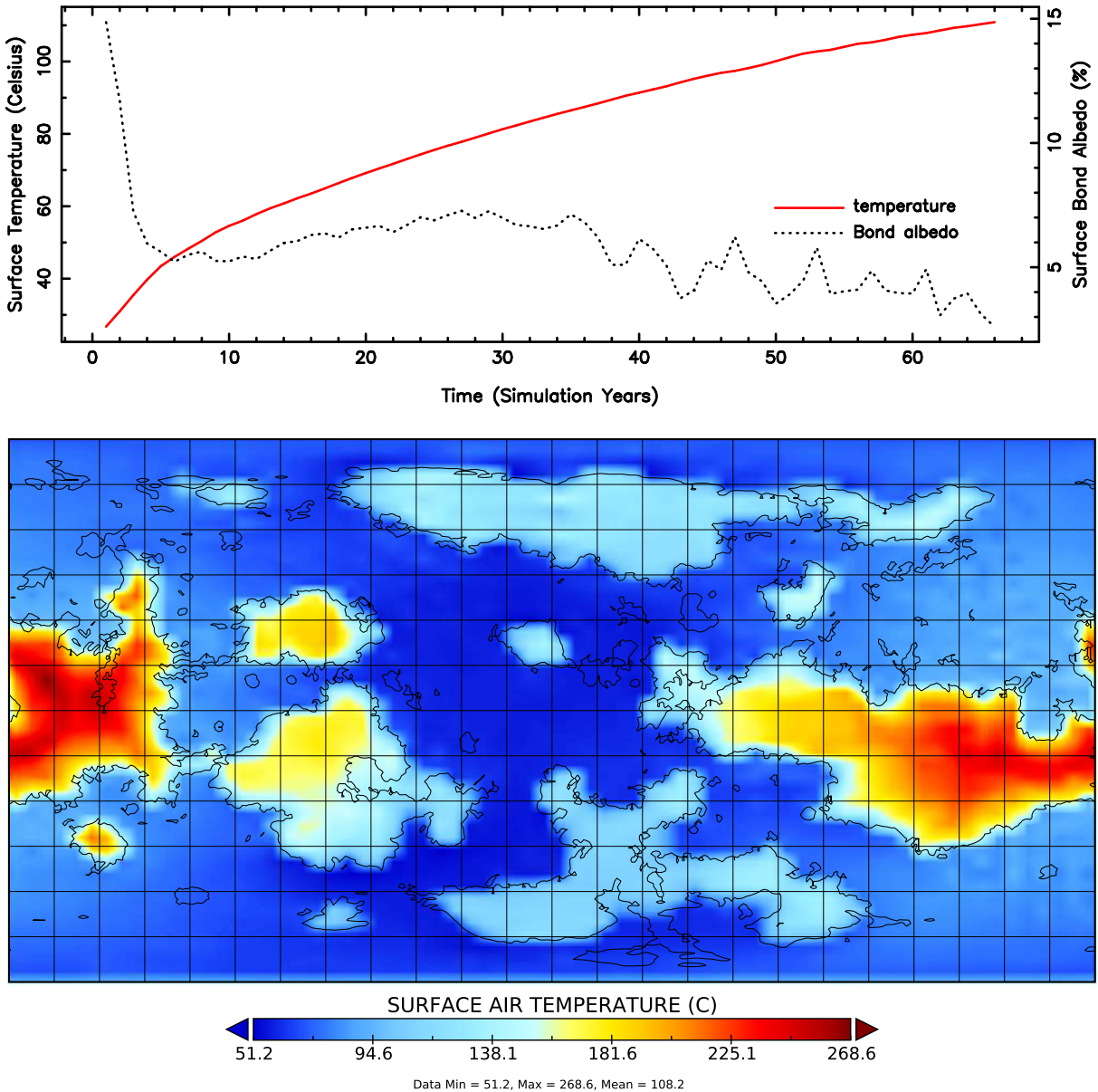}
  \end{center}
  \caption{Results from ROCKE-3D Simulation 2 (see Table~\ref{simtab})
    that utilizes the orbital period and incident flux from
    \citet{ang17}. We adopt a 400/1 CO$_2$/CH$_4$ atmosphere and a
    Venus topography with a liquid water ocean. Top panel: the change
    in surface temperature (red solid line) and Bond albedo (dotted
    black line) of the surface as a function of simulation years
    (orbits). The mean surface temperature at the end of the
    simulation is 108.2\degr~C. Bottom panel: heat map of the
    planetary surface at the conclusion of the simulation with the
    Venus topography overlay. The initial jump in Bond albedo (less so
    for temperature) is due to the fact that the model starts out with
    modern Earth initial conditions and it takes time for the model to
    adjust to the higher insolation being received. See Appendix B
    ``Equilibrium” in \citet{way18} for more details regarding how
    ROCKE-3D comes into net radiative balance, or what we term
    temperature ``equilibrium.” }
  \label{sim2fig}
\end{figure*}

\begin{figure*}
  \begin{center}
    \includegraphics[angle=270,width=16.0cm]{f03a.ps} \vspace{0.4cm}\\
    \includegraphics[clip,width=16.0cm]{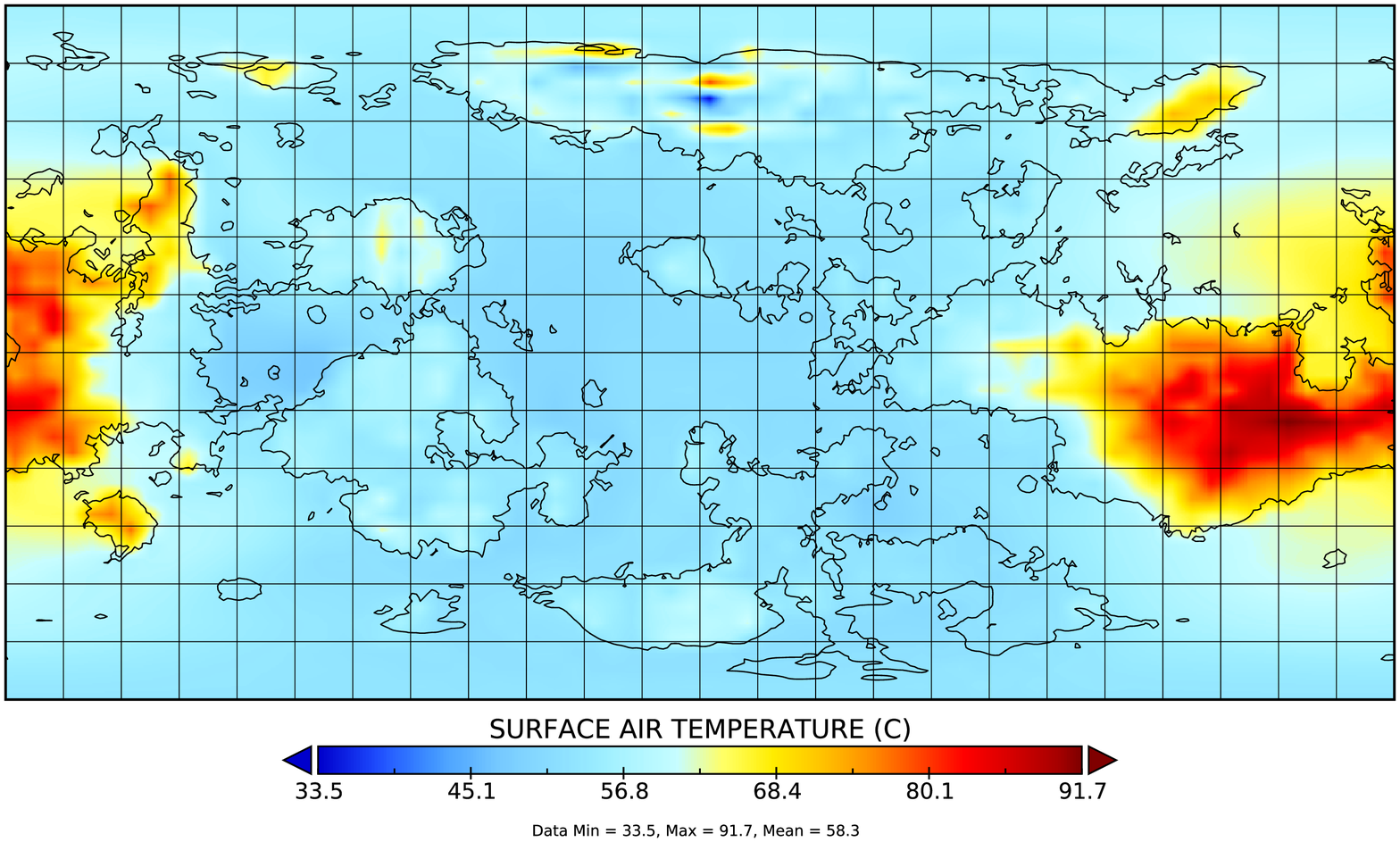}
  \end{center}
  \caption{Results from ROCKE-3D Simulation 9 (see Table~\ref{simtab})
    that utilizes the orbital period and incident flux from
    \citet{ang17}.  We adopt a 400/1 CO$_2$/CH$_4$ atmosphere and a
    Venus topography with a liquid water ocean. Top panel: the change
    in surface temperature (red solid line) and Bond albedo (dotted
    black line) of the surface as a function of simulation years
    (orbits). Bottom panel: heat map of the planetary surface at the
    conclusion of the simulation with the Venus topography overlay. As
    in Figure \ref{sim2fig} the large initial jump in Bond albedo and
    temperature is due to the fact that the model starts out with
    modern Earth initial conditions. The mean surface temperature at
    the end of the simulation is 58.3\degr~C. Note that the length of
    the simulation and surface air temperature limits are distinct
    from those in Figure \ref{sim2fig}.}
  \label{sim9fig}
\end{figure*}

Our strategy with the simulations in Table~\ref{simtab} was to see
whether ROCKE-3D might reach a stable state as we decreased the
insolation, adjusted the orbital period, or changed the
topography. ROCKE-3D is not capable of accurately modeling the
present-day climate of Kepler-1649b, but we demonstrate where the
model reaches stable conditions without exceeding the valid ROCKE-3D
boundaries (as described above) for a similar system with lower
insolation and a longer orbital period. We describe the simulations in
detail below.

Simulation 1. Present-day estimates of Kepler-1649b's insolation and
orbital parameters with an aquaplanet topography. The model ran for
300 orbits before errors in the radiative heating fluxes became too
great for the radiative transfer scheme and the model crashed.

Simulation 2. The same as Simulation 1, but changed from an aquaplanet
to a paleo-Venus-like topography (see Way16) with a $\sim$900m deep
bathtub ocean. Different topographies were attempted (not shown
herein) because Way16 showed that topography can make a large
difference in the climate dynamics and hence surface temperatures. The
temperatures in Simulation 2 were similar to those in Simulation
1. The model completed only 66 orbits before crashing for the same
reasons as Simulation 1. Figure~\ref{sim2fig} contains the evolution
of the surface temperature and Bond albedo as a function of the orbit
number (top panel), and a heat map of the surface temperature with a
Venus topography overlay (bottom panel).

The rapid rise in surface temperature (Figure~\ref{sim2fig}, top
panel) causes substantial surface H$_2$O to transfer to the lower
atmosphere, and even to the stratosphere. This H$_2$O transfer process
occurs for all of the simulations to varying degrees, as seen in the
last two columns of Table~\ref{simtab}. As long as surface reservoirs
of H$_2$O exist (lakes and/or oceans) the lower atmosphere will always
have more water vapor than the upper atmosphere as it takes time for
convection to transport water upward in the atmosphere. The low Bond
albedo is probably due to the fact that the only clouds present are at
the eastern terminator. It is otherwise a cloud-free sky and this
tends to reduce the albedo, as seen in planetary albedo maps produced
from the simulation.

Simulation 3. The same as Simulation 2, but with CO$_{2}$=100ppmv and
CH$_{4}$=0 to test how a change in major greenhouse gases would affect
the surface temperatures. No improvement was apparent and the model
crashed in the radiation after 116 orbits.

Simulations 4--7. The same as Simulation 1, but with insolation
lowered from 2.0 times that of present-day Earth to 1.4. As the
insolation was lowered, the model completed more orbits while getting
closer to equilibrium, yet ultimately still crashed in the radiation.

Simulation 8. The insolation was lowered to present-day Earth (1367 W
m$^{-2}$) and utilized the same shallow ocean Venus land/sea mask as
in Way16. The model completed more orbits and the radiative balance
was far closer to our ideal of $\pm$0.2 W m$^{-2}$. In this case, the
high latitudes began to experience quite low temperatures. ROCKE-3D's
ocean component crashed when a shallow-ocean grid cell at high
latitude froze to the bottom\footnote{This is an unfortunate
  ``feature" of ROCKE-3D that the maintainers are working to
  fix.}. However, unlike the previous simulations, this simulation has
the most accurate climate state prediction given its input parameters
(columns 2--6 in Table~\ref{simtab}) as it is closer to radiative
balance than the others.

Simulation 9. To compare with one of the paleo-Venus simulations in
Way16 (see Simulation D), we lowered the orbital period to 16 Earth
days in length and set the insolation to the same value Venus would
have experienced at 2.9 Gya ($F_p=1.47$). In the Way16 case, the mean
surface temperature reached 56$^{\circ}$C, while the max was
84$^{\circ}$C. In Simulation 9 the mean and max were slightly higher
at values of 58.3$^{\circ}$C and 91.7$^{\circ}$C. Unlike Simulation D
in Way16, this simulation crashed as the surface temperatures climbed
to values outside the bounds of the ROCKE-3D radiation tables. Figure
\ref{sim9fig} demonstrates that the surface temperature of Simulation
9 was possibly beginning to stabilize along with the albedo. So this
simulation may not be far from equilibrium. Regardless, there are two
probable causes for the differences between Simulation D in Way16 and
Simulation 9. First, each simulation uses a very different solar
spectrum. The Kepler-1649b spectrum is heavily weighted toward the
infrared, unlike that of our own sun at 2.9~Gya.  There are a number
of large water vapor absorption bands in the infrared and these play a
role in the increased temperatures (leading to greater heating in the
atmosphere and a lower ocean/lake albedo at the surface).  Second,
while the orbital period of Simulation 9 is 16 Earth days in length,
it is not exactly the same as the 16 day long rotation period of
Simulation D in Way16. Simulation 9 is tidally locked, and hence the
far side of the planet remains in perpetual darkness, whereas
Simulation D is not. This will have some influence on the climate
dynamics of the model as shown in a suite of ROCKE-3D simulations in
\citet{way18} Figure 1. The tidally locked world of Simulation 9 will
fall more into the regime of the slow rotators of \citet{way18} Figure
1 (those with rotation periods of 64 days and longer). There is a
transition from circulation that transports heat poleward (rotation
periods less 64 days) to a circulation of rising motion on the day
side and sinking motion on the night side where transport from day to
night occurs over the poles and at the terminator. \citet{yan13} has
also shown this to be true for tidally locked planets, as is the case
for Simulation 9.

Simulation 10. This was a continuation of Simulation 7 with the same
insolation, but to see whether the climate dynamics would change
substantially if the orbital period was significantly slowed to 50
Earth days in length.  Indeed it had a markedly strong effect and this
simulation is essentially in equilibrium. Here we see the effect of
the single equator-to-pole Hadley cells and the subsequent large cloud
bank at the substellar point that increases the planetary albedo and
shields the planet from the high insolation of the host star, as first
shown in \cite{yang2014} and explicitly demonstrated for a slowly
rotating ancient Venus in Way16.

There are two major caveats to consider when examining the simulations
of Table 1. First, the second-to-last column demonstrates the fact
that, with the exception of Simulation 8, all of these simulations are
well within the moist-greenhouse limit as defined by
\citet{kas88,kas93}. Values of the stratospheric water vapor mixing
ratio $f$(H$_2$O) greater than $3 \times 10^{−3}$ H$_2$O to air (kg)
push one to this moist-greenhouse limit and most simulations are well
above that limit. This implies, according to \cite{kas93} that the
timescale for the loss of all of Earth's ocean water is less than the
age of the present-day Earth.

Secondly, we have to consider whether the specific humidity at any
atmospheric level is above $\sim$10\%. This would indicate that
H$_{2}$O has become a non-negligible part of the atmospheric mass and
errors in the dynamics will begin to increase markedly (see Section
2.1 in \citealt{way18}). Normally the highest values of specific
humidity are found at the surface and indeed, with the exception of
Simulations 8 and 9, they are all well above 10\%.



\section{Transmission Spectra and Detectability}
\label{spectra}

The transmission spectrum of Venus has been observed and simulated on
numerous occasions. Analysis of transmission spectra from the 2004
transit of Venus by \citet{hed11} resulted in the detection of CO$_2$
absorption and was discussed within the context of
exoplanets. Similarly, the 2012 transit of Venus afforded an
additional opportunity to study a high signal-to-noise transmission
spectrum and compare with synthetic spectra of the Venusian atmosphere
\citep{gar12}. Near-infrared observations of Venus have detected water
vapor in the Venusian troposphere \citep{cha13}. The role of
transmission spectra in characterizing exoplanets was discussed by
\citet{bar16}, emphasizing the effect of clouds in producing ambiguous
results. Further simulation of Venusian transmission spectra by
\citet{ehr12} identified absorption features that originated above the
cloud deck that may be used as Venus analog identifiers for
exoplanets.

\begin{figure*}
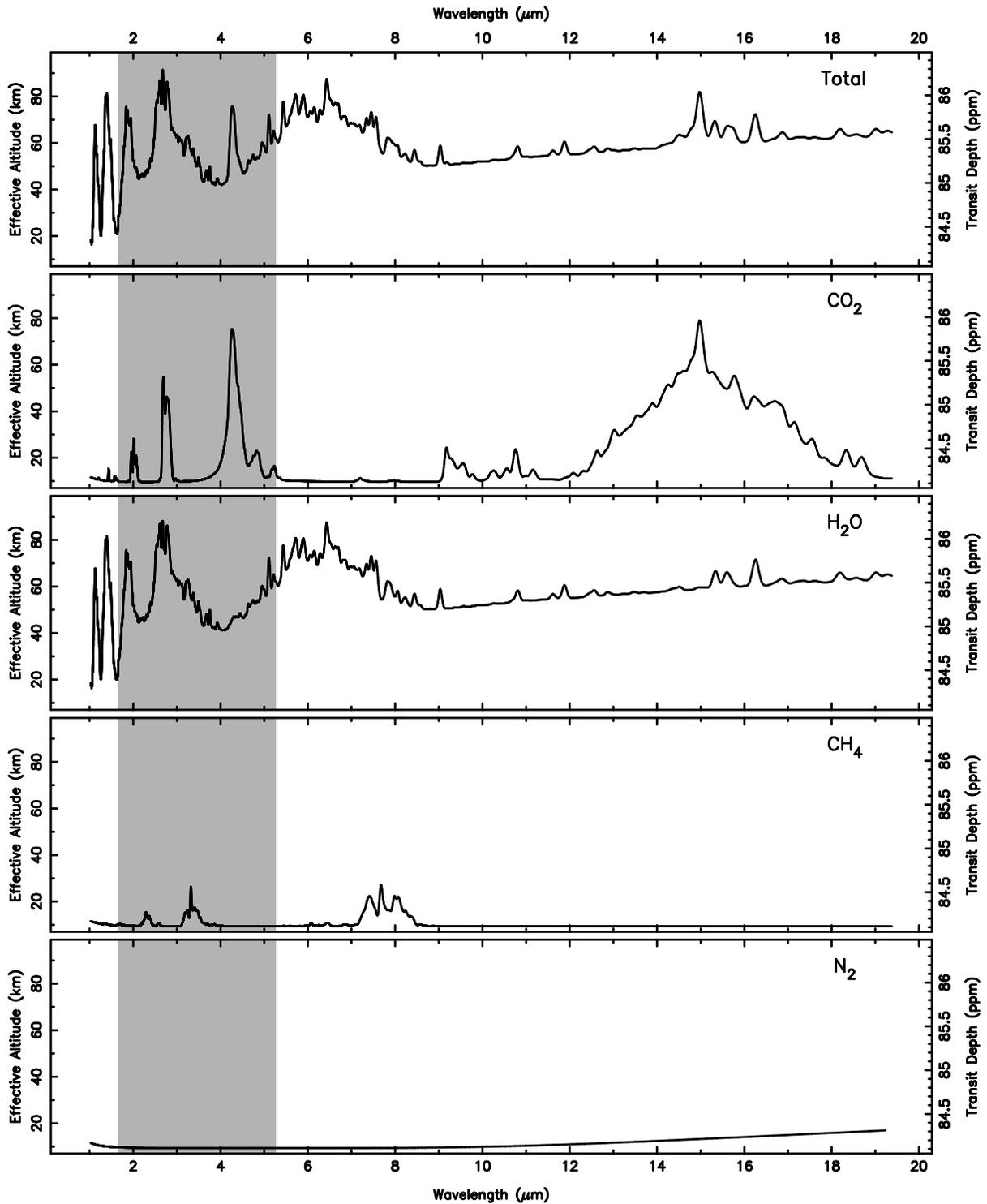

  \begin{center}
    \includegraphics[angle=270,width=16.0cm]{f04a.ps} \\
    \includegraphics[angle=270,width=16.0cm]{f04b.ps} \\
    \includegraphics[angle=270,width=16.0cm]{f04c.ps} \\
    \includegraphics[angle=270,width=16.0cm]{f04d.ps} \\
    \includegraphics[angle=270,width=16.0cm]{f04e.ps}
  \end{center}
  \caption{Synthetic 1--20 micron transmission spectra of the
    Kepler-1649b atmosphere using the results of Simulation 2 (see
    Section~\ref{climate} and Table~\ref{simtab}). The top panel is
    the total transmission spectrum including all molecular
    species. The lower panels, from top to bottom, show the
    contributions of CO$_2$, H$_2$O, CH$_4$, and N$_2$
    respectively. The shaded region shows the extent of the
    JWST/NIRSpec passband.}
  \label{specfig}
\end{figure*}

We utilize a publicly available
code\footnote{https://github.com/GyonShojaah/mock\_observation\_modelE.git}
that was designed to produce transmission spectra from ROCKE-3D output
files. The code reads the vertical profiles at the terminator and
computes the transmission spectrum, probing each grid cell separately,
and accounting for Rayleigh scattering, refraction, and molecular
absorption based on HITRAN2012 \citep{rot13}. We applied this code to
Simulation 2 (as described in Section~\ref{climate}) since the
incident flux for the simulation most closely represents the
measurements provided by \citet{ang17} and the Venus topography allows
direct comparison with early Venus. The simulated 1--20~$\mu$m
transmission spectrum based on the outputs of Simulation 2 is shown in
the top panel of Figure~\ref{specfig}. The figure also shows the
spectrum contributions from the individual components of CO$_2$,
H$_2$O, CH$_4$, and N$_2$. As described in Section~\ref{climate} and
shown in the top panel of Figure~\ref{sim2fig}, the rapid rise in
surface temperature results in the transfer of H$_2$O from the oceans
to the atmosphere. This is why the total spectrum shown in
Figure~\ref{specfig} is dominated by water vapor. The CO$_2$
atmospheric content produces substantial absorption features in the
1.5--5.5~$\mu$m range as well as strong absorption effects centered at
$\sim$15~$\mu$m. The contributions of both CH$_4$ and N$_2$ are
negligible apart from small absorption features from CH$_4$ between 1
and 9~$\mu$m and from a slightly positive N$_2$ gradient from smaller
to longer wavelengths. An important aspect of this transmission
spectrum is that it represents the spectrum of a potential transition
state of the atmosphere since the climate simulation did not achieve
radiative balance. Though the simulation ended before radiative
balance was achieved, the clear rise in temperature incidates that the
final state of the atmosphere is a runaway greenhouse.

Future prospects are promising for detecting key atmospheric
signatures in exoplanets that can help distinguish between Earth and
Venus analogs. JWST will revolutionize the field of exoplanet
atmospheres by virtue of nearly continuous, long-baseline
observations, broader wavelength coverage, and higher spectral
resolution than existing ground and space-based facilities
\citep{gre16}. While HST's Wide-Field Camera 3 (0.8--1.7~$\mu$m) has
enabled transit spectroscopy atmospheric characterization for dozens
of planets, it is primarily sensitive to water features
\citep{Deming2013}, and the interpretation of spectra is highly
degenerate and model-dependent \citep{bea18}. With broader wavelength
coverage and high spectral resolution, JWST will be able to probe a
much wider range of chemical species with fewer model assumptions,
including the components shown in Figure~\ref{specfig} (CO$_2$,
H$_2$O, CH$_4$, and N$_2$).

JWST is equipped with four visible to mid-IR instruments (NIRCam,
NIRISS, NIRSpec, and MIRI) that span 0.6--28~$\mu$m. JWST will have an
Early Release Science (ERS) program focused on transiting exoplanets
to explore these instruments and exercise their various modes in order
to demonstrate their capabilities and characterize their systematics
\citep{bea18}. The ERS Panchromatic Transmission Program will use
NIRISS (0.6--2.8~$\mu$m), NIRCam (2.4--4.0~$\mu$m), and NIRSpec
(1.66--5.27~$\mu$m; see Figure~\ref{specfig}) to obtain a panchromatic
near-IR transmission spectrum of a single planet to calibrate the
instruments and establish the best strategies for obtaining transit
spectra for future cycles. This program will be particularly important
for probing the atmospheres of small exoplanets. This will require
knowledge of the noise and systematics and ultimately determine the
number of transit observations required to build sufficient precision
to detect atmospheric features.

Several studies have investigated the optimal observing strategies for
detecting molecular species of small planets orbiting low-mass stars,
and have found consistent signal-to-noise requirements of about 10
transits \citep{Morley2017,Louie2018,bat18}. These requirements are
dependent on both stellar (distance, magnitude, size) and planetary
(size, mass) parameters. For the nearby (12 pc) star TRAPPIST-1, which
is 18\% the mass of the sun, five transit observations are planned
with the NIRSpec Prism mode for two of its Earth-sized planets in the
HZ (under the JWST Guaranteed Time Observations program). While
Kepler-1649b is a prime Venus analog in terms of size and incident
flux, the star is comparable in size to TRAPPIST-1 but resides about
67~pc from the sun and is thus relatively faint (Kepler magnitude
of 17), which will make it a challenge for follow-up even for
JWST. {\it TESS} is expected to discover dozens of small planets,
including Venus analogs like Kepler-1649b, that orbit nearby bright
M-dwarfs \citep{ric15,Barclay2018}. Many of these planets will be
amenable to mass measurements and will be excellent targets for JWST
\citep{kem18} and future Venus analog studies.


\section{Discussion}
\label{discussion}

There are numerous proposed causes which can lead atmospheric
evolution into a runaway greenhouse state. These include an increase
in stellar luminosity \citep{lec13}, an increase in greenhouse gases
\citep{gol13}, tidal heating for planets in eccentric orbits
\citep{bar13}, and a dependence on the latitudinal surface water
distribution \citep{kod18}. The simulations presented in this paper
primarily focus on the atmospheric evolution after initial conditions
that explore a range of atmospheric compositions and surface water
initial states. The transition from a surface liquid water state to a
full runaway greenhouse is not easily modeled by any 3D climate model,
but our simulations do show the transition into a moist greenhouse. It
is unclear how long such transitional atmospheric states persist and
the synthetic transmission spectra produced for these states, such as
the example described in Section~\ref{spectra}, may help identify
planets which are undergoing such transitions. Note that the
transition to a moist greenhouse need not guarantee that a runaway
greenhouse will be the final outcome. Indeed it was pointed out by
\citet{wor13} that a moist greenhouse triggered relatively late in the
evolution of the host star such that atmospheric erosion and water
loss due to XUV radiation is minimized, could result in a
CO$_2$/H$_2$O-rich atmosphere that retains surface liquid water.

The follow-up of suitable potential Venus analog targets will require
a concerted effort that will utilize the capabilities of such
facilities as JWST. There are numerous teams producing simulations
using observation times required to fully characterize exoplanet
atmospheric signatures, such as the observing strategies proposed by
\citet{bat18} and the prioritization framework described by
\citet{kem18}. As noted earlier, the relative faintness of Kepler-1649
($J = 13.379$) presents a significant challenge in achieving the
needed signal-to-noise ratio (S/N) for atmospheric
characterization. Equation~1 of \citet{kem18} describes a transmission
spectroscopy metric that approximates the S/N for a 10~hr
observation with JWST/NIRISS, where \citet{kem18} recommend a S/N~$>
10$ for terrestrail targets. Using the measured properties of the
Kepler-1649 system and the methodology of \citet{kem18}, we calculate
a transmission spectrum S/N of $\sim$2.1. For a brighter star of $J =
10$ ($d = 14.1$~pcs), the S/N is $\sim$10.2, and for $J = 8$ ($d =
5.6$~pcs), the S/N is $\sim$25.6. Therefore, discoveries of similar
systems that are much closer to our Sun, such as the discoveries from
the {\it TESS} mission, will present viable targets to more fully
investigate the characteristics of potential runaway greenhouse
atmospheres.


\section{Conclusions}
\label{conclusions}

The celestial sphere has been subjected to intensive monitoring from
transit surveys, both from the ground and from space. The photometric
precision of these surveys combined with the continuous coverage
provided by space-based facilities has changed the shape of the known
exoplanet demographics and revealed numerous terrestrial
exoplanets. The bias of the transit method toward shorter period
planets means that terrestrial planets with incident fluxes
substantially higher than that received by the Earth comprise the bulk
of the current and near-future discovery space. Hence, in the near
term, the majority of the terrestrial exoplanet inventory will reside
in a radiation environment that more closely matches that of Venus
than of the Earth. This presents a significant challenge since the
atmosphere, surface, and interior of Venus have many fundamental
questions that have yet to be answered \citep{tay18}. The vast number
of terrestrial planet discoveries in the VZ of their host stars makes
it imperative that stellar astronomers collaborate with planetary
scientists to understand Venusian environments and correctly interpret
exoplanet data.

We have presented here a detailed study of the potential Venus analog
Kepler-1649b using the ROCKE-3D climate package and adopting a range
of starting conditions. Our simulations show that the planetary
surface environment is highly likely to exist in a runaway greenhouse
state since all of our simulations show a rapid and significant rise
in surface temperature when adopting the measured values for the
system \citep{kas88}. These simulations then validate the initial
speculation by \citet{ang17} that the planet may indeed be a Venus
analog. Our synthetic transmission spectrum for the planetary
atmosphere will provide a useful basis for interpreting data acquired
for similar planets orbiting brighter host stars that are accessible
via future space-based instrumentation. The identification of such
Venus analogs will provide invaluable insight into the diversity of
exoplanet demographics, evolution of planetary atmospheres, the
conditions for runaway greenhouses, and the inner boundaries of
habitable planetary surface environments.


\section*{Acknowledgements}

The authors would like to thank Sonny Harman and Yuka Fujii for useful
discussions regarding this work. E.V.Q. and M.J.W. are grateful for
support from GSFC Sellers Exoplanet Environments Collaboration (SEEC).
This research has made use of the Habitable Zone Gallery at
hzgallery.org. The results reported herein benefited from
collaborations and/or information exchange within NASA's Nexus for
Exoplanet System Science (NExSS) research coordination network
sponsored by NASA's Science Mission Directorate.


\end{document}